\documentstyle[preprint,tighten,prb,aps,eqsecnum]{revtex}
\draft
\title{Interacting Electrons on a Square Fermi Surface}
\author{A. Luther}
\address{NORDITA, Blegdamsvej 17, DK-2100 Copenhagen \O, Denmark}
\date{\today}
\begin{document}
\maketitle
\begin{abstract}
Electronic states near a square Fermi surface are mapped onto quantum chains.
Using boson-fermion duality on the chains, the bosonic part  of the interaction
is isolated and diagonalized. These interactions destroy  Fermi liquid
behavior. Non-boson interactions are also generated by this mapping, and  give
rise to a new perturbation theory about the boson problem. A case with strong
repulsions  between parallel faces  is studied and solved. There is spin-charge
separation and the square Fermi surface remains square under doping. At
half-filling, there is a charge gap and  insulating behavior together with
gapless spin excitations. This mapping appears to be a general tool for
understanding the properties of interacting electrons on a square Fermi
surface.
\end{abstract}

\narrowtext
\section{INTRODUCTION}

\noindent
The problem of interacting electrons in two space dimensions is thought to be
central to an understanding of high temperature superconductivity\cite{alan1}.
Ordering in three dimensions, according to this scenario, arises from the
coupling of highly correlated layers\cite{alan2}. Attention naturally focusses
on the properties of the layers and the possibility of unconventional behavior
in two dimensions.\\
\\
A standard model often studied in this context is the Hubbard model. This
begins with electrons which may tunnel between nearest neighbor sites on a
square lattice, with amplitude $t$, and adds an on-site repulsion of strength
$U$. In comparison with the corresponding problem in one dimension\cite{alan3},
relatively little is known about this model.\\
\\
Perturbation theory in $U$ shows very interesting and complicated behavior.
With the starting point of an average occupation number of one electron per
site, serious divergences are encountered\cite{alan4}. These have been analyzed
for more general interactions than the Hubbard model interaction\cite{alan5}
and are worse than the usual logarithmic type found in other problems.
Evidently, there is a need for new techniques to be developed.\\
\\
The starting point of one electron per site has a square Fermi surface. This
paper discusses a novel mapping of the square Fermi surface onto one
dimensional chains.  Two sets of chains arise, one for each axis of the square.
Interactions can be decomposed into couplings between parallel faces of the
square and adjacent faces.  These, in turn, result in interactions between the
chains.  Boson-fermion duality\cite{alan6} can be used on each chain separately
to find the equivalent boson problem, giving a precise definition of this
duality for the square Fermi surface.\\
\\
In a natural way, the interactions separate into purely bosonic terms plus
other new operators.  The bosonic terms can be diagonalized exactly and produce
behavior similar to the Luttinger model in the sense that there are no fermion
quasi-particles or Fermi liquid behavior\cite{alan7}.  There is complete
separation of the charge and spin degrees of freedom, as previously found in
purely one dimensional models\cite{alan8}. As a consequence, charge excitations
have no spin and pure spin excitations no charge.\\
\\
Non-bosonic terms produced by this mapping give rise to a new starting point
for perturbation theory. This begins with the free electron kinetic energy plus
the bosonic terms taken together as the unperturbed Hamiltonian. The non-boson
terms arising in the mapping are then taken as the perturbation. It is then
important to analyze the operators in this new perturbation theory, to see if
they are relevant or irrelevant, with respect to the bosonic starting point.\\
\\
This technique of separating interactions into a bosonic piece plus other
operators is familiar from many other problems, such as the Kondo
model\cite{alan9}, the backward scattering model\cite{alan8}, or the
sine-Gordon problem in 1D\cite{alan10}. It is slightly more complicated here
due to the mapping of the square Fermi surface onto chains.\\
\\
One of the complicating features of the square Fermi surface concerns the
modulation of the Fermi velocity along the face --- in fact it vanishes at the
corners. In this new perturbation analysis, it turns out that the operator
causing modulation is irrelevant for any interaction. The large distance
behavior is characterized by a constant velocity.\\
\\
It might be argued that doping would destroy these results, since the Fermi
surface should then depart from a square. That is only correct for nearly  free
electrons. For sufficiently strong interactions, the operator which causes
deviations from the square becomes irrelevant. A square remains a square under
doping, only the value of the Fermi momentum changes.\\
\\
Generally, it is found that two different types of operators are relevant for
repulsive interactions. These correspond to Umklapp terms and an
antiferromagnetic spin-flip process. For a restricted case of repulsive
interactions between parallel faces, these extra terms can also be
diagonalized. They lead to a gap in the charge excitation spectrum and,
therefore, insulating behavior. There are gapless spin excitations. This range
of coupling constants also exhibits spin-charge separation.\\
\\
There have been studies of coupled 1D chains using models for the
coupling\cite{alan11} which differ from that encountered in this paper.
Consequently, the results cannot be applied here.  Other studies of the square
Fermi surface problem have concentrated on models with boson-like
interactions\cite{alan12} and found non-Fermi liquid behavior\cite{alan13}.  It
is not clear how to extend these to include the non-boson interactions which
are found to be relevant for short range interactions.\\
\\
The techniques introduced in this paper are generally applicable for
understanding the relevant degrees of freedom for interacting electrons on a
square Fermi surface. Although the special case of interactions between
parallel faces is emphasized, it should be easy to generalize these results to
other models, including the Hubbard model. Future work will doubtlessly explore
the fall possibilities of this mapping. Some interesting directions are
indicated in the discussion section.

\section{MAPPING THE SQUARE FERMI SURFACE ONTO CHAINS}

\noindent
The general problem of interacting electrons in two space dimensions, described
by the tight binding approximation on square  lattices, simplifies if attention
is focused on low energy excitations near the Fermi surface. For the
half-filled hand, the Fermi surface is a square, and the important excitations
preserve this symmetry. Intuitively, excitations perpendicular to a face of the
square can be thought of as one-dimensional in character if the momentum
parallel
to the face is zero. For large parallel momentum, this picture is not obvious
since the Fermi velocity can vary with position along the face. In addition,
low energy excitations between different faces can also exist, confusing a
separation into excitations of each separate face.\\
\\
These different excitations are contained in the kinetic energy operator of the
tight-binding model,
\begin{equation}
\displaystyle{H_o = - t \sum_{\langle ij\rangle\sigma} ~a^\dagger_{i\sigma}
a_{j\sigma} ~ + \mu\sum_{i\sigma}~a^\dagger_{i\sigma} a_{i\sigma}}
\label{eq:201}
\end{equation}

\noindent
where $a_{i\sigma}$ is the Fermi operator of an electron at site $i$ with spin
$\sigma~, t$ is the hopping matrix element, and the sum over $i$ and $j$ is
restricted to nearest neighbors on a square lattice. After a Fourier
transformation, and rotation by 45$^{o}$, this can be written as:
\begin{equation}
\displaystyle{H_o = -4t\sum_{\vec{k}\sigma} ~\cos (k_x s^\prime) ~\cos (k_y
s^\prime )~a^\dagger_{\vec{k},\sigma} ~ a_{\vec{k},\sigma}}
\label{eq:202}
\end{equation}

\noindent
where $s^\prime = s/\sqrt{2}$. The fermion field operator is defined before
rotation as
$N^{-1}\sum_{\vec{k}^\prime}a_{\vec{k}^\prime,\sigma}e^{i\vec{k}^\prime
\cdot\vec{R}^\prime}$, for a lattice of $N^2$ sites, area $L^2$, lattice
constant ``s'', and $\vec{k}^\prime = 2\pi L^{-1} (n^\prime_2 , n^\prime_y)$.
The sum is the over over the Brillouin zone, and the chemical potential $\mu$,
in
Eq.\ (\ref{eq:201}) is chosen such that the energy of Eq.\ (\ref{eq:202})
vanishes for $k_x$ or $k_y$ equal to $\pm$ $k_F$, with $k_F = \pi(\sqrt{2}
s)^{-1}$. This defines the
square Fermi surface.\\
\\
Consider now that face for $k_x$ near $k_F$. The energy can be linearized to
find
\begin{equation}
\displaystyle{E_1 (k_x) = v_F (k_y) ~ (k_x - k_F)}
\label{eq:203}
\end{equation}

\noindent
where $v_F (k_y) = (2 \sqrt{2} ts)\cos (k_y s^\prime)$ and the ``1'' subscript
refers to the particular face with $k_x = k_F$,
$-k_F < k_y \leq k_F$, and positive (or zero) Fermi velocity. There is a
similar linearization for the other three faces, labelled by ``2'', with $k_x =
-k_F$, ``3'' with $k_y = k_F$, and ``4'' with $k_y = -k_F$, in all cases with
positive (or zero) Fermi velocity.\\
\\
For a given spin, the Fermi operator describing the ``1'' face is given, after
rotation, by
\begin{equation}
\displaystyle{\psi^\prime_1 (x,y) = \frac{1}{N}\sum_{k_y , k_x} ~a_{1 k}
{}~e^{\displaystyle{i k_y y + i k_x x}}}
\label{eq:204}
\end{equation}

\noindent
The $\vec{k}$ momenta are 45$^{o}$ rotations of the $\vec{k}^\prime$, $k_y$
runs from $-k_F$ to $k_F$, $x$ and $y$ are rotated coordinates. With
$\vec{R}^\prime = s(m,n)$, these are given by $ \vec{R} = (x,y) $ with $
\sqrt{2}x = s(m+n)$ and $\sqrt{2} y = s(n-m)$. The equal time correlation
function $\langle\psi^{\prime\dagger}_1 (x,y) \psi_1^\prime\rangle$  is given
by
\widetext
\begin{equation}
\begin{array}{lll}
\displaystyle{\langle\psi^{\prime\dagger}_1 (x,y) \psi_1^\prime \rangle
}&=&\displaystyle{
 s^2\int_{-k_F}^{k_F}\frac{dk_y}{2\pi} } ~e^{\displaystyle{-i k_y y }}
\int_{-\infty}^{k_F}  ~ \frac{d k_x}{2\pi}  ~e^{\displaystyle{-i k_x x-\alpha |
k_x - k_F|}}\\
&&\\
&=&\pi k_F^{-1}  g(l-l^\prime)~
\left( \frac{\displaystyle ie^{-i k_F x}}
{\displaystyle 2\pi(x + i\alpha)}\right)
\end{array}
\label{eq:205}
\end{equation}
\narrowtext

\noindent
where $g(l) = 2^{-1} \delta_{l,o} + (\pi l)^{-1} \sin (\pi l 2^{-1})$, with $l
= n-m$ representing a discrete ``y'' coordinate, and $\alpha$ is a cutoff of
order $k_F^{-1}$ for states far away from the Fermi surface. In these formula
the sums over $k$ have been replaced by integrals, using the invariance of
$dk_x~dk_y$ under rotation. \\
\\
This result is a product of a function of the $y$ variable times a purely 1D
expectation value. It suggests that the Fermi operator describing excitations
near the ``1'' face can be written in the form:
\begin{equation}
\displaystyle{\psi^\prime_1 (x,y) = \sqrt{\frac{\pi}{k_F}}~\sum^{2N}_{l^\prime}
g(l-l^\prime) \psi_{1l^\prime} (x)}
\label{eq:206}
\end{equation}

\noindent
where the prime denotes the operator for the square Fermi surface,
the operator on the right of Eq.\ (\ref{eq:206})
describes fermions on independent
chains, and $\sqrt{2} y = sl$. The integer $l$ runs from $1$ to $2N$, since
$l = n-m$ with $m$ and $n$ running from $1$ to $N$.\\
\\
The expectation value for independent chains\\
$\displaystyle{\langle\psi^\dagger_{1l_1} (x) \psi_{1 l_2})\rangle}$
with normalization $\psi_l (x) =$\\$ L^{-\frac{1}{2}} ~ \sum_{k_x} ~ a_l (k_x)
{}~ e^{i k_x x}$, is:    
\begin{equation}
\begin{array}{c}
\displaystyle{\langle\psi^\dagger_{1l_1} (x) \psi_{1l_2}\rangle}
\displaystyle{\delta_{l_1 l_2} L^{-1} \sum_{k<k_F}
{}~e^{-i k x}}\\
\\
= \displaystyle{\delta_{l_1 , l_2}
{}~\left(\frac{i e^{\displaystyle{-i k_F x}}}{2 \pi (x+ i \alpha)}\right)}
\end{array}
\label{eq:207}
\end{equation}

\noindent
which gives Eq.\ (\ref{eq:205})
when $\sum_{l_1} g (l-l_1) g ( l_1 - l^\prime) = g(l-l^\prime)$ is used with
Eq.\ (\ref{eq:206}).\\
\\
It is also interesting to consider the mapping of Eq.\ (\ref{eq:206})
in the language of Fourier transforms. For $2N$ independent chains, the
transformed operator is
\begin{equation}
\displaystyle{\psi_{1 n_y} (x) = (2N)^{-1/2} \sum^{2N}_{l=1} \psi_{1l} (x) ~
e^{i p_y l}}
\end{equation}

\noindent
with $p_y = 2\pi n_y (2N)^{-1} $ and $n_y$ in  the interval $-N$ to $+N$. This
corresponds to the interval in $k_y$ from $-2k_F$ to $+2k_F$, since $2k_F y =
\pi l$. The discretization of this interval is consistent with the spacing of
lines of equal $k_y$, which is
$2\pi L^{-1} 2^{-\frac{1}{2}}$, and the phase factor $k_y y = n_y (2 \pi L^{-1}
2^{-\frac{1}{2}} )y = p_y l$. The point with $k_y = k_F$ has $n_y =
\frac{N}{2}$, and the square
Fermi surface thus corresponds to $2N$ independent $1D$ chains with the Fourier
transform momentum restricted to the half-interval.\\
\\
An obvious question with this mapping onto independent chains concerns the
continuum limit used in the normalization of Eq.\ (\ref{eq:207}),
contrasted to that of Eq.\ (\ref{eq:204}). They differ by $\sqrt{s}$, which is
to be expected for a purely 1D system\cite{alan14}. For the discrete lattice
system in 1D, the free fermion expectation value $\langle\psi^\dagger_1 (m)
\psi_1 (o)\rangle$  behaves as $m^{-1}$ for large $m$, while
$\langle\psi^\dagger_1 (x) \psi_1 (o)\rangle$  for the continuum model behaves
as $x^{-1}$. Since $x\rightarrow ms$, the operators must differ by the
multiplicative factor of $\sqrt{s}$. With the convention used in Eq.\
(\ref{eq:207}),
the correlation functions are correctly reproduced by Eq.\ (\ref{eq:206}),
and that is used in this paper.\\
\\
This continuum limit cannot be applied to the other dimension, since the Fermi
surface degeneracy occurs over a finite range in momentum space. The continuum
limit can only be taken for states perpendicular to the Fermi surface, while
the other direction remains discrete.\\
\\
Similar considerations apply to the construction of the Fermi operator for
$k_x$ near $-k_F$, the ``2'' states. The corresponding expectation value is
found to be:
\begin{equation}
\displaystyle{\langle\psi^\prime_{2}~ ^{\dagger}(x,y) \psi_2^\prime \rangle  =
\pi k_F^{-1}~ g(l-l^\prime) ~
\frac{i e^{i k_F x}}{2\pi (-x+i \alpha)}}
\label{eq:209}
\end{equation}

\noindent
and the $\psi^\prime_2$ field is represented by
\begin{equation}
\psi^\prime_2 (x,y) = \pi k_F^{-1}~\sum^{2N}_{l^\prime = 1} g(l - l^\prime)
\psi_{2l^\prime} (x)
\label{eq:210}
\end{equation}

\noindent
where $\psi_{2 l}(x)$ is the operator for the ``2'' fermion on chain $l$.\\
\\
The operator content of this mapping is clear. With the Fourier transform
operator on the Fermi surface defined by
\begin{displaymath}
\begin{array}{lllll}
\psi^\prime_{1 n_y} (x) &=& \displaystyle{\psi_{1 n_y} (x),} &-\frac{N}{2} \leq
&n_y \leq \frac{N}{2}
\end{array}
\end{displaymath}
and:
\begin{equation}
\begin{array}{lllll}
\psi_{1 n_y} (x) &=& \displaystyle{(2N)^{-\frac{1}{2}}  \sum^{2N}_{l=1}
\psi_{1l} (x)~
e^{-i p_y l}} &-N < & n_y < N
\end{array}
\label{eq:211}
\end{equation}

\noindent
It is seen that the Fermi surface operator is just the chain operator
restricted to half the transverse Brillouin zone. This mapping is useful
because interactions between the various faces of the Fermi surface have simple
representations when expressed in terms of the chain picture. This is discussed
in the following section.\\
\\
Coupling between the chains could arise from the dependence of the Fermi
velocity on the momentum along the face. This coupling is contained in the
$k_y$ Fourier transform of Eq.\ (\ref{eq:202}) and Eq.\ (\ref{eq:203}):
\begin{equation}
\displaystyle{H_1 = \sum^{N/2}_{n_y = -\frac{N}{2}}~\int^L_o \frac{dx}{i} ~ v_F
(p_y) \psi^{\prime\dagger}_{1 n_y} ~ (x) \frac{\partial}{\partial
x}\psi^\prime_{1 n_y} (x)}
\label{eq:212}
\end{equation}

\noindent
Since the sum in this equation is restricted to the half interval, the $\psi_1$
operators are equal to the unprimed operators, and the inverse
Fourier transform of Eq.\ (\ref{eq:204}) can be used to find
\begin{equation}
H_1 = \sum_{l,l^\prime} ~\int ~\frac{dx}{i}~  t_{l , l^\prime}
{}~\psi^\dagger_{1l} (x)  \frac{\partial}{\partial x}~\psi_{1l^\prime} (x)
\label{eq:213}
\end{equation}

\noindent
where
\begin{eqnarray*}
t_{l,l^\prime} = (2N)^{-1} \sum^{\frac{N}{2}}_{n_y = -\frac{N}{2}} ~ v_F (p_y)
e^{i p_y (l - l^\prime)}
\end{eqnarray*}

\noindent
In the following section, it is shown that the off-diagonal elements of
$t_{l,l^\prime}$ are irrelevant operators when interactions are included. The
diagonal element represents an asymptotic Fermi velocity, $v_o$, of magnitude
\begin{equation}
v_o = \frac{(\sqrt{2} ts)}{\pi}
\label{eq:214}
\end{equation}

\noindent
and $t_{l,l^\prime} \rightarrow v_o \delta_{l,l^\prime}$.\\
\\
Keeping only the diagonal element, the kinetic energy is a sum of independent
chains. Each chain can be separately bosonized, giving the result:
\begin{equation}
H_1 = \pi  v_o \sum^{2N}_{l=1} \int^L_o ~dx ~ \rho_{1l} (x) \rho_{1l} (x)
\label{eq:215}
\end{equation}

\noindent
where $\rho_{1l} (x) = \psi_{1l}^\dagger (x) \psi_{1l} (x)$, and the product is
normal ordered. The Fourier transforms of $\rho_{1l} (x) $ are:
\begin{equation}
\begin{array}{lll}
\displaystyle{\rho_1 (\vec{n})} &=& \displaystyle{\sum^{2N}_{l=1}~ \int_o^L dx
{}~ e^{i p_y l + i k_x x}}~ \rho_{1l} (x)\\
&&\\
\displaystyle{\rho_{1l} (x) }&=& \displaystyle{(2NL)^{-1}~\sum_{n_y n_x}~ e^{-i
k_x x - i p_y l}~ \rho_1  (\vec{n})}
\end{array}
\label{eq:216}
\end{equation}

\noindent
With these relations, the kinetic energy for face ``1'' can be written as:
\begin{equation}
\displaystyle{H_1 = \frac{\pi v_o}{2NL }~ \sum_{\vec{n}} \rho_1 (\vec{n})
\rho_1 (-\vec{n})}
\label{eq:217}
\end{equation}

\noindent
where $n_y$ runs from $-N$ to $N$ and the sums over $n_x$ are to be cutoff as
is the 1D case.\\
\\
\noindent
The commutator of $\psi^\prime_{1l} (x)$, from Eq.\ (\ref{eq:206}),
with the kinetic energy is
\begin{equation}
[\psi_{1l} (x) , H_1 ] = i v_o \frac{\partial}{\partial x} \psi_{1l} (x)
\label{eq:218}
\end{equation}

\noindent
using either Eq.\ (\ref{eq:212}) or Eq.\ (\ref{eq:215})
for $H_1$, as a result of boson-fermion for the 1D system on chain $l$.

\noindent
These steps can be repeated for the other faces of the Fermi surface, including
spin, giving the total kinetic operator (under the assumption that $t_{l ,
l^\prime} \rightarrow v_o \delta_{l , l^\prime}$) as:

\begin{equation}
H_o = \frac{\pi v_o}{2 NL}~ \sum_{\vec{n} , \sigma , \beta} \rho_{\beta\sigma}
(\vec{n})  \rho_{\beta\sigma} (-\vec{n})
\label{eq:219}
\end{equation}

\noindent
where the products are normal ordered, $\sigma = \pm 1$ is a spin index, and
$\beta$ = 1 to 4 corresponding to the four faces.

\noindent
The commutation relations are given by

\begin{equation}
\begin{array}{lllll}
\displaystyle{ \left[\rho_{1\sigma^\prime} (\vec{n}^\prime), \rho_{1\sigma}
(\vec{n}) \right] } &=&
\displaystyle{- \left[\rho_{2\sigma} (\vec{n}^\prime), \rho_{2\sigma} (\vec{n})
\right] } &=&
\displaystyle{2 N n_x \delta_{\vec{n} , - \vec{n}^\prime} \delta_{\sigma ,
\sigma^\prime} }\\
&&&&\\
\displaystyle{ \left[\rho_{3\sigma^\prime} (\vec{n}^\prime), \rho_{3\sigma}
(\vec{n}) \right] }&=&
\displaystyle{- \left[\rho_{4\sigma^\prime} (\vec{n}^\prime), \rho_{4\sigma}
(\vec{n})  \right]} &=&
\displaystyle{ 2 N n_y \delta_{\vec{n} , - \vec{n}^\prime} \delta_{\sigma ,
\sigma^\prime} }
\end{array}
\label{eq:220}
\end{equation}

\noindent
With all other commutations vanishing. Fermi operators for the various faces
are:
\begin{equation}
\begin{array}{lll}
\psi^\prime_{\beta , \sigma} (x,y) &= &\displaystyle{ \sqrt{\frac{\pi}{k_F}}
{}~\sum^{2N}_{l^\prime = 1} ~g(l-l^\prime) \psi_{l^\prime , \beta , \sigma}
(x)}\\
&&\\
\psi^\prime_{\beta^\prime , \sigma} (x,y) &= & \displaystyle{
\sqrt{\frac{\pi}{k_F}} ~\sum^{2N}_{k^\prime = 1} ~g(k-k^\prime) \psi_{k^\prime
, \beta^\prime , \sigma} (y)}
\end{array}
\label{eq:221}
\end{equation}

\noindent
where $\sigma$ is the spin index, $\beta = 1,2$ the face index, $l$ is the
discrete coordinate corresponding to $y$, $\sqrt{2}y = s l$, and $x$ the
continuum variable on chain $l^\prime$. The other two faces are rotated
versions of the above, with $\beta^\prime = 3,4$ the face index, $k$ the
discrete $x$ coordinate, and $y$ the continuum variable on chain $k^\prime$.\\
\\

\noindent
The necessity to construct fermion operators unsymmetrically in $x$ and $y$
follows from the degeneracies over finite regions in momentum space, the flat
pieces of the Fermi surface. The mapping onto independent chains gives the
correct multiple point fermion expectation values, since it preserves the
correct operator structure. This construction therefore gives the correct
asymptotic behavior of these correlation functions and the dynamics near the
Fermi surface. It is particularly useful when interactions between the
different faces are included as described in the following section. It should
be emphasized that no approximation has been made beyond linearization of the
free electron spectrum at the Fermi surface and the assumption that
$t_{ll^\prime}$ is diagonal, as discussed in Section IV.

\section{DENSITY OPERATORS FOR THE SQUARE FERMI SURFACE}

\noindent
The interactions between electron are often described by the Hubbard model,
\begin{equation}
\displaystyle{H = H_o + U \sum_i n_{i +} n_{i -}}
\label{eq:301}
\end{equation}

\noindent
where $H_o$ is given by Eq.\ (\ref{eq:201}), $U$ is the on-site repulsion
between electrons of opposite spins, and $n_{i\sigma}$ the number operator at
site $i$ with spin $\sigma$. The previous section presented a mapping of $H_o$
onto independent chains, and it is natural to understand the interaction term
in this representation. Basic to this construction is the assumption that
states near the Fermi surface are important, since these are correctly
reproduced by that mapping.\\
\\
In principle, the procedure to follow is straightforward. An electron operator
is just the sum of the four operators, $\psi = \psi_1 + \psi_2 + \psi_3 +
\psi_4$, from the four faces. Each can be mapped into independent chains, with
two orthogonal  sets of chains for the two axes of the square. Since  $n_{i+}
= \psi^\dagger_+ ~ \psi_+$, there are 256 terms, but there are also great
simplifications due to symmetry and operator dimension. \\
\\
The first simplification occurs in the density operators themselves. Consider
the contribution to
the spin $(+1)$ density from face ``1'', using the mapping of
Eq.\ (\ref{eq:219})
\begin{eqnarray}
\psi^{\prime\dagger}_{1+} (x,y) \psi^\prime_{1+} (x,y) =
\nonumber\\
\displaystyle{ \pi k_F^{-1} \sum_{l_1 l_2} g(l-l_1) g(l - l_2) \psi^\dagger_{1
l_1 +} (x) \psi_{1 l_2 +} (x)    }
\label{eq:302}
\end{eqnarray}
%\begin{equation}
%\displaystyle{\psi^{\prime\dagger}_{1+} (x,y) \psi^\prime_{1+} (x,y) = \pi
%%k_F^{-1} \sum_{l_1 l_2} g(l-l_1) g(l - l_2) \psi^\dagger_{1 l_1 +} (x)
%%\psi_{1 l_2 +} (x)    }
%\end{equation}

\noindent
The terms with $l_1 = l_2$ gives a purely bosonic contribution, since
$\psi^\dagger_{1l_1 +} (x)  \psi_{1l_1 +} (x) = \rho_{1 l_1 +} (x) $ is of the
form discussed in
Eq.\ (\ref{eq:216}).
Defining this boson piece by $\rho_{1+}^\prime (x,y)$, results in:
\begin{equation}
\displaystyle{\rho^\prime_{1+} (x,y) = \pi k_F^{-1} \sum_{l_1} g^2 (l-l_1)
\rho_{1 l_1 +} (x)}
\label{eq:303}
\end{equation}

\noindent
and has the Fourier transform, with conventions from Eq.\ (\ref{eq:216}),
given by
$\pi k_F^{-1} ~ \rho^\prime_{1+} (\vec{n})$, with
\begin{equation}
\rho^\prime_{1+} (\vec{n}) = f(n_y) \rho_{1+} (\vec{n})
\label{eq:304}
\end{equation}

\noindent
In this equation,  $f(n_y)$ is the Fourier transform of $g^2 (l)$,
\begin{eqnarray*}
f(n_y) = \sum_{l=1}^{2N} ~ e^{i p_y l} g^2 (l) = \frac{1}{2} \left(1 -
\frac{|n_y|}{N}\right) = \frac{1}{2} \left( 1 - \frac{|p_y|}{\pi}\right)
\end{eqnarray*}
for $|n_y| \leq N, ~ (|p_y|\leq \pi)$. Using the commutation relations of the
$\rho_{\alpha, \sigma} (\vec{n})$ operators on the right hand
side of Eq.\ (\ref{eq:304}),
given in Eq.\ (\ref{eq:217}),
the operator algebra for the boson part of the Fermi
surface density operator can be found.\\
\\
\noindent
It is interesting to study the equal-time density-density correlation function,
for free particles, and determine how much is contained in the purely bosonic
piece of Eq.\ (\ref{eq:304}).
The correlation function is found using Eq.\ (\ref{eq:206}),
\begin{eqnarray}
&\displaystyle{\langle\psi^{\prime\dagger}_{1} (x,y) \psi_1^\prime (x,y)
\psi^{\prime\dagger}_1 \psi^\prime_1\rangle}\nonumber\\
=& \displaystyle{\langle\psi^{\prime}_1 (x,y) \psi_1^\prime\rangle\langle
\psi^{\prime}_1 (x,y) \psi^{\prime\dagger}_1\rangle}\nonumber\\
&\nonumber\\
=& \displaystyle{\pi^2 k_F^{-2} g^2 (l)
\left(\frac{i}{2\pi}~\frac{1}{x+i\alpha}\right)^2}
\label{eq:305}
\end{eqnarray}

\noindent
while the boson part of the density operator contributes:
\begin{equation}
\displaystyle{\langle\rho^\prime_1 (x,y) \rho^\prime_1 \rangle  = \pi^2
k_F^{-2} \sum_{l^\prime} g^2 (l-l^\prime) g^2 (l^\prime)
\left( \frac{i}{2\pi} ~\frac{1}{x + i \alpha}\right)^{2} }
\label{eq:306}
\end{equation}

\noindent
The sum over $l^\prime$ gives $(2 \pi^2 l^2)^{-1} + (4)^{-1} \delta_{l, o}$,
with $g^2(l)  $ from Eq.\ (\ref{eq:205}).
The difference arises from the restriction to equal chains contained in
$\rho^\prime_1 (x,y)$. In general, contributions form unequal chains must
be examined, and if relevant, they should be retained. Nonetheless, a
purely bosonic piece can always be extracted from a density operator.
The unequal chain terms are analyzed in the next section.\\
\\
The boson contribution to the Hubbard model interaction of
Eq.\ (\ref{eq:301}) can be extracted using this procedure. After the notation
used in Eq.\ (\ref{eq:202}),
the contribution from the ``1'' face is given by
\begin{equation}
\begin{array}{lll}
U\sum_i n_{i+} n_{i-} &\rightarrow &\displaystyle{U \int \frac{dx}{s} \sum_l
\rho^\prime_{1 l +} (x) \rho^\prime_{1 l -} (x)}\\
&&\\
&=& \displaystyle{\frac{U}{s} \pi^2 k_F^{-2}   \frac{1}{2NL} \sum_{\vec{n}} f^2
(n_y) \rho_{1+} (\vec{n}) \rho_{1-} (-\vec{n}) \cdot}\\
&&\\
&=& \displaystyle{\frac{(Us)}{NL}\sum_{\vec{n}} f^2 (n_y) \rho_{1+} (\vec{n})
\rho_{1-} (-\vec{n})}
\end{array}
\label{eq:307}
\end{equation}

\noindent
Note that the combination $Us$ enters here, and this defines the
coupling constant for the continuum limit, $Us$. In the solutions discussed
below, only the ratio $Us/v_o$ appears, and from
Eq.\ (\ref{eq:212}), the lattice constant cancels, as in the 1D
Hubbard model continuum limit. An interaction which exists only at one site can
be generalized to have different interactions strengths between different
faces. By symmetry, one expects three different interactions, equal faces,
opposite faces, and adjacent faces. There is also a complete separation of
charge and spin degrees of freedom in these interactions, giving six coupling
 constants in all.\\
\\
Charge-spin separation follows from choosing linear combination in the form
\begin{equation}
\begin{array}{ll}
&\displaystyle{\sqrt{2} \rho_\alpha (\vec{n}) = \rho_{\alpha +} (\vec{n}) +
\rho_{\alpha-} (\vec{n})}\\
&\\
&\displaystyle{\sqrt{2} \sigma_\alpha (\vec{n}) = \rho_{\alpha +} (\vec{n}) -
\rho_{\alpha-} (\vec{n})}
\end{array}
\label{eq:308}
\end{equation}

\noindent
where $\alpha$ is the face index. These operators satisfy
\begin{eqnarray*}
[\sigma_{\alpha^\prime}  (\vec{n}^\prime ) , \sigma_{\alpha}  (\vec{n} ) ] =
[\rho_{\alpha^\prime} (\vec{n}^\prime ), \rho_{\alpha} (\vec{n}) ] =
\delta_{\alpha , \alpha^\prime} \delta_{\vec{n} - \vec{n}^\prime} ~ 2N (
\hat{\epsilon}_\alpha \cdot \vec{n})
\end{eqnarray*}
where $\hat{\epsilon}_1 = \hat{x} , \hat{\epsilon}_2 = -\hat{x} ,
\hat{\epsilon}_3 = + \hat{y}$, and $\hat{\epsilon}_4 = -\hat{y}$, and all other
commutators vanish. An interaction term of the form in Eq.\ (\ref{eq:307})
can be re-written, using
\begin{eqnarray}
\displaystyle{2 \sum_{\vec{n},\alpha\beta} f_{\alpha\beta}
(\vec{n})  \rho_{\alpha +} (\vec{n}) \rho_{\beta -}
(-\vec{n})}\nonumber\\
\displaystyle{ = \sum_{\vec{n},\alpha\beta}  f_{\alpha\beta}
(\vec{n}) \left[  \rho_{\alpha} (\vec{n}) \rho_{\beta}
(-\vec{n}) - \sigma_\alpha (\vec{n}) \sigma_\beta (-\vec{n})\right]
  }
\label{eq:309}
\end{eqnarray}

\noindent
provided $f_{\alpha \beta} $ is even in $\vec{n}$ and symmetric under
interchange
of $\alpha$ and $\beta$, which is the case here.\\
\\
The most general interaction allowed by this construction is then
\begin{equation}
\displaystyle{H_B = \frac{1}{NL}\sum_{\vec{n}, \alpha \beta}\left[
\rho^\prime_\alpha (\vec{n}) A_{\alpha \beta} \rho^\prime_\beta (-\vec{n}) -
\sigma^\prime_\alpha (\vec{n})
B_{\alpha\beta} \sigma^\prime_\beta (-\vec{n})\right]}
\label{eq:310}
\end{equation}

\noindent
with $A_{\alpha\alpha} = V_1$ and $B_{\alpha\alpha} = U_1$ representing
scattering on the same face; $A_{12} = A_{21} = A_{34} = A_{43} = V_2$ and
$B_{12} = B_{21} = B_{34} = B_{43} = U_2$ for opposite faces, while all other
elements of $A_{\alpha \beta} = V_3$ and $B_{\alpha \beta} = U_3$, representing
scattering from adjacent faces, are equal. In a similar fashion, the boson form
of the kinetic energy,
Eq.\ (\ref{eq:219}), can be separated into spin and charge operators,
using Eq.\ (\ref{eq:308}). The result is
\begin{equation}
H_o = \frac{\pi v_o}{NL}\sum_{\vec{n}\alpha}\left[\rho_\alpha  (\vec{n})
\rho_\alpha (-\vec{n}) + \sigma_\alpha (\vec{n}) \sigma_\alpha (-\vec{n}\right]
\label{eq:311}
\end{equation}

\noindent
and together with Eq.\ (\ref{eq:310}),
provides a complete description of the bosonic degree of freedom for the square
Fermi surface. This provides a starting point from which the relevant (or
irrelevant) operators are to be determined. \\
\\
For the Hubbard model, all of the six coupling constants are equal, with
magnitude $2Us$. This equality is not obviously useful for they may renormalize
in different ways when relevant operators, coming from non-boson terms in the
interaction, are included. Although the boson Hamiltonian $H_o + H_B$ from
Eq.\ (\ref{eq:310}) and Eq.\ (\ref{eq:311})
can be diagonalized, the problem is simplified somewhat if coupling between
adjacent faces is omitted, $V_3 = U_3 = 0$, the procedure followed below.\\
\\
The parameters describing boson interactions between parallel faces can be more
intuitively understood by inverting the spin-charge separation. In this way,
$H_B$ for parallel face interactions can also be written in the form:
\widetext
\begin{equation}
\begin{array}{lll}
H_B &=& \displaystyle{\frac{1}{2NL}\sum_{\vec{n}\alpha \sigma \sigma^\prime}
f^2} (n_y) \rho_{\alpha\sigma} (\vec{n}) \left[ (V_1 - U_1) \delta_{\sigma ,
\sigma^\prime} +
(V_1 + U_1) \delta_{\sigma , -\sigma^\prime} \right] \rho_{\alpha \sigma^\prime
    (-\vec{n})}\\
&&\\
&&\displaystyle{+ \frac{1}{NL}\sum_{\vec{n}\sigma , \sigma^\prime} f^2 (n_y)
\rho_{1\sigma}
(\vec{n}) \left[ (V_2 - U_2) \delta_{\sigma , \sigma^\prime} +
(V_2 + U_2 ) \delta_{\sigma , -\sigma^\prime} \right] \rho_{2\sigma^\prime}
(-\vec{n}) }\\
&&\\
&&\displaystyle{+(3-4 ~\mbox{terms}})
\end{array}
\label{eq:312}
\end{equation}
\narrowtext

\noindent
where $\alpha = 1,2$. The contribution from the ``3'' and ``4'' faces is found
by replacing $f(n_y)$ by $f(n_x)$, $\rho_1$ by $\rho_3$, and $\rho_2$ by
$\rho_4$.\\
\\
All of these are forward scattering processes. The first term has small
momentum transfer on the same face, while the second describes small momentum
transfer on opposite faces. Although this representation makes a physical
interpretation more evident, solutions are more conveniently found using the
separation into spin and charge operators.\\
\\
It should also be recognized that the generalized model with six coupling
constants is not the most general, even for the boson degrees of freedom. In
order to have these unequal, it is necessary to have a momentum dependent
interaction, rather than the on-site interaction of the Hubbard model. As a
result, the momentum dependence along the square faces would also appear, which
changes the momentum dependence contained in the $f(n)$ factor appearing in
Eq.\ (\ref{eq:304}). The model which retains only $V_1 , V_2 , U_{2}$ and
$U_2$, discussed in the following section, not only provides insight into the
more general problem, but perhaps can serve as a phenomenological description
as well.

\section{RELEVANT OPERATORS AND THE BOSON PROBLEM}

\noindent
The preceding section discussed the construction of boson operators which are
contained in the fermion density operators, and showed how to extract products
of boson operators from the Hubbard interaction. In this section, the boson
problem is diagonalized, and the remaining terms contained in the interaction,
are isolated and discussed.\\
\\
Central issues concern the value of the single electron states and the possible
breakdown of Fermi liquid theory. The conditions for spin-charge separation
need to be analyzed, along with the nature of the excitation spectrum. Further
questions involve the consequences of the modulation of the Fermi velocity
along the face and the stability of the square Fermi surface to doping away
from half-filling. These problems can all be studied using the boson
Hamiltonian, from the proceeding section.\\
\\
The starting point for the boson problem is $H_o + H_B$, given by
Eq.\ (\ref{eq:310}) and Eq.\ (\ref{eq:311}), which can be written as:
\widetext
\begin{equation}
\begin{array}{lll}
H_o + H_B &=& \displaystyle{\frac{\pi v_o}{NL}\sum_{\vec{n} , \alpha}
\left\{ \left(1 + V_1 \frac{f^2_\alpha (\vec{n})}{\pi v_o} \right)
\rho_\alpha  (\vec{n}) \rho_\alpha (-\vec{n}) +
\left( 1 - U_1 \frac{f^2_\alpha  (\vec{n})}{\pi v_o}\right)
\sigma_\alpha  (\vec{n}) \sigma_\alpha (\vec{n}) \right.}\\
&&\\
&&\displaystyle{ +  \frac{2V_2}{\pi v_o} f^2 (n_y) \rho_1 (n) \rho_2 (-n) -
\frac{2 U_2}{\pi v_o} f^2 (n_y) \sigma_1 (\vec{n}) \sigma_2 (-   (\vec{n})
}\\
&&\\
&&\displaystyle{\left.+  \frac{2V_2}{\pi v_o} f^2_3 (n_x) \rho_3 (n) \rho_3
(-\vec{n}) -
\frac{2 U_2}{\pi v_o} f^2 (n_x) \sigma_4 (\vec{n}) \sigma_4 (-\vec{n})
\right\}}\\
\end{array}
\label{eq:401}
\end{equation}
\narrowtext

\noindent
Here the sum over $\vec{n}$ is defined from $-N$ to $N$ for momentum parallel
to the face in question, with an exponential cutoff perpendicular to the face,
as in the 1D problem. The function $f_\alpha (\vec{n})$ equals $f(n_y)$ for
$\alpha = 1,2$ and $f(n_x)$ for $\alpha = 3,4$. Note that the coupling constant
enters as the ratio $V_1 / v_o$ etc.,  and the lattice constant contained in
these objects cancels. Terms containing a zero in the denominator are to be
omitted. They
corresponds to zero modes, containing no space-time dependence, and must be
treated separately.\\
\\
Separation into spin and charge operators is explicit, and each sector can be
diagonalized separately, using canonical transformations such as:
\begin{equation}
\begin{array}{lll}
\displaystyle{\rho^\prime_1 (\vec{n})}&=&\displaystyle{\cosh \eta (n_y) \rho_1
(\vec{n}) - \sinh \eta (n_y) \rho_2 (\vec{n}) }\\
&&\\
\displaystyle{\rho^\prime_2 (\vec{n})}&=& \displaystyle{\cosh \eta (n_y) \rho_2
(\vec{n})   - \sinh \eta (n_y) \rho_1 (\vec{n}) }\\
\end{array}
\label{eq:402}
\end{equation}

\noindent
generated by $\rho_1^\prime (\vec{n}) = e^{S_1} \rho_1 (\vec{n}) e^{-S_1}$
with
 $S_1 = (2N)^{-1} \sum_{\vec{n}} n^{-1}_x \eta (n_y) \rho_1 (\vec{n}) \rho_2
(-\vec{n}).$ A similar transformation diagonalizes the spin degrees of freedom
for the 1-2 faces, with $S_2 =  (2N^{-1}) \sum_{\vec{n}} n^{-1}_x \mu (n_y)
\sigma_1 (\vec{n}) \sigma_2 (-\vec{n})$. The transformation for the 3-4 faces
are found by interchanging $n_x$ and $n_y$, with  1$\rightarrow$3 and
2$\rightarrow$4 for the boson subscripts.\\
\\
Applying these transformations to Eq.\ (\ref{eq:401}) leads to the diagonal
form
\begin{equation}
\displaystyle{H_D = \frac{\pi v_o}{NL} \sum_{\vec{n} , \alpha}
\left[\gamma_\alpha  (\vec{n})      \rho_\alpha   (\vec{n})  \rho_\alpha
(-\vec{n})  +
\delta_\alpha  (\vec{n})  \sigma_\alpha  (\vec{n})  \sigma_\alpha
(-\vec{n})\right]}
\label{eq:403}
\end{equation}

\noindent
where $\alpha = 1,2,3,4$. The charge velocity renormalizations are given by:
\begin{equation}
\gamma^2_1 (\vec{n}) = \gamma^2_2 (\vec{n}) =
\left(1 + \frac{V_1}{\pi v_o} f^2 (n_y )\right)^2 -
\left(\frac{V_2 f^2 (n_y)}{\pi v_o}\right)^2
\label{eq:404}
\end{equation}

\noindent
with $\gamma_3 (\vec{n}) = \gamma_4 (\vec{n})$ found from Eq.\ (\ref{eq:404})
by replacing $n_y$ by $n_x$. The spin velocities, $\delta_\alpha (\vec{n})$,
are given by $\gamma_\alpha (\vec{n})$ with the replacements $V_1 \rightarrow -
U_1$ and $V_2 \rightarrow - U_2$. In general, the spin and charge velocities
are different, but preserve the $x-y$ symmetry.\\
\\
The parameters used in these diagonalization transformations $S_1$ to $S_4$ are
given by
\begin{equation}
\displaystyle{e^{-4 \eta  (n_y)} =
\frac{1 + \left(\displaystyle{\frac{V_1 - V_2}{\pi v_o}}\right) f^2 (n_y) }{1 +
 \left(\displaystyle{\frac{V_1 + V_2}{\pi v_o}}\right) f^2 (n_y) }}
\label{eq:405}
\end{equation}

\noindent
and $\eta (n_x)$ is Eq.\ (\ref{eq:405}), replacing $n_y$ by $n_x$. For the spin
degrees of freedom, $\mu (n_y)$ is given by Eq.\ (\ref{eq:405}) with $V_1$ and
$V_2$ replaced by
$-U_1$ and $-U_2$ respectively, while $\mu (n_x)$ replaces $n_y$ by $n_x$.\\
\\
This completes the diagonalization of the boson problem. It is seen that the
spectrum is linear in momentum perpendicular to each face, and each will
contribute a term linear in temperature to the specific heat. Interactions
renormalize this linear term, as in the Luttinger model.\\
\\
These transformations can be used to calculate the fermion correlation
function, and determine the relevant operators. It is helpful to use the
boson-fermion duality on each chain separately, representing for example the
fermion operator of Eq.\ (\ref{eq:206}) by
\begin{equation}
\begin{array}{lll}
\psi_{1l\sigma} (x) &=& Z_l ~ e^{\displaystyle{-\varphi_{1l\sigma} (x) + i  k_F
 x}}\\
&&\\
\psi_{2l\sigma} (x) &=& Z_l ~ e^{\displaystyle{\varphi_{2l\sigma}  (x) - i  k_F
x}}
\end{array}
\label{eq:406}
\end{equation}

\noindent
where the ``phase'' fields are defined by
\begin{equation}
\begin{array}{llll}
\displaystyle{\varphi_{\alpha l \sigma} (x) }&=&  \displaystyle{(2N)^{-1}
}\displaystyle{ \sum_{\vec{n}} n_x^{-1} \rho_{\alpha\sigma} (\vec{n})}
{}~e^{\displaystyle{-i k_x x -i p_y l} }\\
&&\\
&=& \displaystyle{2^{-3/2}~ N^{-1}} \displaystyle{\sum_{\vec{n}} n_x^{-1}}
\displaystyle{\left[\rho_{\alpha} (\vec{n}) + \sigma \sigma_{\alpha}
(\vec{n})\right]} ~e^{\displaystyle{-i k_x x - i p_y l} }
\end{array}
\label{eq:407}
\end{equation}

\noindent
with $\alpha = 1,2$, $\sigma = \pm 1$, and separation into charge and spin
operators has been used. A renormalization constant $Z_l$ has been introduced,
which contains the contributions from zero modes corresponding to the $n_x = 0$
term in Eq.\ (\ref{eq:407}). For the purely 1D case\cite{alan6}, $Z = (2
\pi\alpha )^{-\frac{1}{2}}$, but for chains, this can also contain a phase
factor $\exp (i \phi_{ \alpha\sigma} (l))$ corresponding to a uniform fermion
phase on different chains.\\
\\
The corresponding operators for the 3-4 faces are defined by
\begin{equation}
\begin{array}{lll}
\psi_{3k\sigma} (y) &=& Z_k e^{\displaystyle{-\varphi_{3k\sigma}(y) + i k_F y
}}\\
&&\\
\psi_{4k\sigma} (y) &=& Z_k e^{\displaystyle{\varphi_{4k\sigma}(y) - i k_F y }}
\end{array}
\end{equation}

\noindent
and these ``phase'' fields are found from Eq.\ (\ref{eq:407}) by the
replacements 1$\rightarrow$3, 2$\rightarrow$4, and $n_x$ by $n_y$. For these
3-4 face operators, $y$ is now the continuum valuable, and $k$ represents the
discrete $x$-variable. Thus the plane wave phase factor $k_x x + p_y l$ in Eq.\
(\ref{eq:407})
is replaced by
$k_y y + p_x k$, for the $\varphi_{\alpha k\sigma} (y)$ fields, for $\alpha =
3,4$.\\
\\
In the following, it will be shown that fermion operator must be paired on
equal chains, otherwise the corresponding terms are irrelevant. That simplifies
the zero mode factors, $Z$, since products of fermion operators on different
chains commute, and potential single fermion ordering terms drop out.\\
\\
The calculation of a fermion correlation function in the boson ground state
separates into products of operators in the spin and charge sectors, as seen by
the example
\widetext
\begin{displaymath}
\begin{array}{lll}
\displaystyle{\langle\psi^\dagger_{1l +} (x,t) \psi_{10 +} \rangle  }&=&
Z^2 \langle  e^{\varphi_{1l+} (xt) } ~ e^{\displaystyle{-\varphi_{10 +}}
\rangle  ~ e^{\displaystyle{-i k_F x}}}\\
&=& Z^2 e^{\displaystyle{-i k_F x}} ~ A_l (x,t) B_l (x,t)
\end{array}
\end{displaymath}
where
\begin{equation}
A_l(x,t) = \left\langle
e^{\displaystyle{N^{-1}2^{-\frac{3}{2}}
                 \sum_{\vec{n}} n^{-1}_x \rho_1 (\vec{n} , t)
                 e^{\displaystyle{-i k_x x - i p_y l}}}}~
e^{\displaystyle{-N^{-1}2^{-\frac{3}{2}}
                 \sum_{\vec{n}} n^{-1}_x \rho_1 (\vec{n})}}
\right\rangle
\end{equation}

\noindent
with time evolution, and expectation values using the $\rho$ part of the
Hamiltonian in Eq.\ (\ref{eq:403}). In identical manner, the $B$ term replaces
$\rho_1$ with $\sigma_1$, and uses the $\sigma$ part of Eq.\ (\ref{eq:403}).\\
\\
Diagonalizing the Hamiltonian mixes $\rho_1$ and $\rho_2$ according to
Eq.\ (\ref{eq:402}), such that $A$ is a product $A_1 A_2$ involving the ``1''
and ``2'' operators
\begin{equation}
A_{1_l} (x,t) = \left\langle
e^{\displaystyle{N^{-1}2^{-\frac{3}{2}}
                 \sum_n n^{-1}_x\cosh\eta(n_y)\rho_1(\vec{n},t)
                 e^{\displaystyle{-i k_x x - i p_y l}}}}~
e^{\displaystyle{-N^{-1}2^{-\frac{3}{2}}
                 \sum_n n_x^{-1}\cosh\eta(n_y)\rho_1(\vec{n})}}
\right\rangle
\end{equation}

\noindent
with $A_2 (x,t) $ given by replacing $\cosh \eta (n_y)$ by $\sinh \eta (n_y)$
and $\rho_1$ by
$\rho_2$.\\
\\
In this expectation value, the time evolution $\rho_1 (\vec{n} ,t) = \rho_1
(\vec{n}) =
e^{i k_x v_o \gamma (n_y) t}$ is to be used. These can be calculated using the
Baker-Hausdorf formula to find
\begin{equation}
\displaystyle{A_{1l} (x,t) = \exp \left\{ -(4N)^{-1} \sum^N_{n_y = - N}
\sum^{\infty }_{n_x = 1} ~ n^{-1}_x ~ \cosh^2 ~ \eta (n_y) ~e^{-\alpha k_x}
 ~(1 - e^{\displaystyle{i p_y l + i k_x \left(x - v_o \gamma (n_y)
t\right)}})\right\} }
\label{eq:411}
\end{equation}
\narrowtext

\noindent
with $A_{2l} (x,t)$ replacing $x$ by $-x$ and $\cosh^2 \eta (n_y)$ by $\sinh^2
\eta (n_y)$ . Converting the sum over $n_x$ to an integral over $k_x$, it is
seen that this integral diverges to $(-\infty)$ at $k_x = 0$, unless $l= 0 $.
Thus, at $t = 0$:
\begin{equation}
A_{1l} (x,t) = \delta_{l ,o} \left(\frac{i\alpha}{x + i\alpha} \right)^
{\frac{\displaystyle 1 + d\rho}{\displaystyle 2}}
\label{eq:412}
\end{equation}

\noindent
where $d_\rho = N^{-1} \int^N_o d n_y \cosh^2 \eta (n_y)$ . The time dependence
is complicated by the dependence of velocity on $n_y$  but as long as it
remains positive, Eq.\ (\ref{eq:411}) provides a calculation of the operator
dimension of $A_1$, since $t$ scales as $x$. Proceeding to the other
expectation values, there results:
\begin{eqnarray*}
\begin{array}{lll}
A_{2l} (x) &=& \delta_{l,0} \left(\displaystyle{\frac{i \alpha}{-x + i
\alpha}}\right)^{\displaystyle{\frac{d_\rho}{2}}}\\
B_{1l} (x) &=& \delta_{l,0}\left(\displaystyle{\frac{i\alpha}{x +
i\alpha}}\right)^{\displaystyle{\frac{1 + d_\sigma}{2}}}\\
B_{2l} (x) &=& \delta_{l,0}\left(\displaystyle{\frac{i\alpha}{-x +
i\alpha}}\right)^{\displaystyle{\frac{ d_\sigma}{2}}}
\end{array}
\end{eqnarray*}

\noindent
and consequently
\begin{equation}
\langle\psi^\dagger_{1 l +} (x) \psi_{1 0 +}\rangle  = \delta_{l,0} ~ e^{-i k_F
x}
\left(\frac{1}{2\pi \alpha}\right) \left(\frac{i \alpha}{x + i \alpha}\right)
\left(\frac{\alpha^2}{x^2 + \alpha^2}\right)^{\displaystyle{\frac{d_\rho +
d_\sigma}{2}}}
\label{eq:413}
\end{equation}

\noindent
where $d_\sigma$ is found as in Eq.\ (\ref{eq:412}) for the spin degrees of
freedom replacing $\eta$ by $\mu$. This is the result of independent chains of
the Luttinger model with a suitably chosen coupling constant. The time
dependence is more complicated since the boson spectrum depends on $n_y$, but
the essential physics remains 1D in character.\\
\\
The occupation number for an electron $\langle n_{1l\sigma}(k_x)\rangle$  on
chain $l$ is an integral\cite{alan7} over the correlation function of
Eq.\ (\ref{eq:413}). It is seen that only fermion operators  on the same chain
contribute, so the occupation number is independent of $l$. Since the $x$
dependence of Eq.\ (\ref{eq:413}) is the same as in the Luttinger model, it can
be concluded that $\langle n_{1l\sigma} (k_x)\rangle$  is continuous at $k_x =
k_F$. For other faces, the result is identical. Bosonic portions of the
interactions on a square Fermi surface destroy Fermi-liquid behaviors.\\
\\
It is now appropriate to check the assumption of Eq.\ (\ref{eq:214}) concerning
the off-diagonal matrix elements of $t_{l,l^\prime}$, using the boson problem
of Eq.\ (\ref{eq:403}) as the starting point. This is done using perturbation
theory in $t_{l,l^\prime}$ to check whether this operator is relevant. One of
the terms to be studied is
\begin{equation}
T_{1\sigma} (x) = \sum_{l\neq l^1} t_{ll^\prime} \psi^\dagger_{1l\sigma} (x)
\frac{\partial}{\partial x} \psi_{1l^\prime \sigma}^{} (x)
\label{eq:414}
\end{equation}

\noindent
and in second order perturbation theory would contribute
\begin{equation}
\int\!dx\int\!dt\int\!dx^\prime\int\!dt^\prime\langle  T_{1\sigma} (x,t)
T_{1\sigma} (x^\prime,t^\prime)\rangle
\label{eq:415}
\end{equation}

\noindent
where time evolution and averaging is with respect to $H_o + H_B$.\\
\\
Using the boson representation for the $\psi_{\alpha l \sigma}(x)$ operator,
the expectation value in Eq.\ (\ref{eq:414}) is calculated to be proportional
to:
\begin{displaymath}
\sum_{l_1 \neq l_2} t^2_{l_1 -l_2}  \frac{1}{(R - V^\prime \tau + i\alpha)^4}
{}~\left(
\frac{\alpha^2}{R^2 - V^{\prime^2}  \tau^2 + \alpha^2}\right)^{d (l_1 - l_2)}
\end{displaymath}

\noindent
where $V^\prime$ is an average (finite) velocity, instead of doing the integral
over $\gamma (n_y), R = x - x^\prime , \tau = t - t^\prime$, and $d(l_1 - l_2)$
is given by:
\widetext
\begin{equation}
d(l_1 - l_2) = N^{-1} \int^N_o d n_y \left[ \sinh^2 \eta (n_y) + \sinh^2 \mu
(n_y)\right] \left[ 1 - \cos p_y (l_1-l_2)\right]
\end{equation}
\narrowtext

\noindent
For large $l_1 - l_2$, the cosine term gives a vanishing contribution, and
Eq.\ (\ref{eq:415}) reduces to the expected result from naive dimensional
analysis, using Eq.\ (\ref{eq:412}), with $d(\infty) = d_\rho + d_\sigma$. \\
\\
Since $d (l_1 - l_2) > 0$, the integrals over space-time in
Eq.\ (\ref{eq:415})
are convergent at large distance giving a factor of the two dimensional volume.
The sum over $l_1$ and $l_2$ is proportional to $N$, since $t_{l_1 - l_2}$
falls off as
$(l_1- l_2)^{-2}$. This result means $T_{1 l_1 \sigma} (x)$  is an irrelevant
perturbation, the usual result in two  space-time dimensions, when the operator
dimension is greater than two. The reduction from two to one space dimensions
is a consequence of the finite sum over chains, a result reminiscent of
independent one dimensional chains.\\
\\
In similar fashion, the contributions from the other operators can be seen to
be irrelevant. Consequently, the modulation of the Fermi velocity along the
face of the Fermi surface is an irrelevant perturbation about the boson
Hamiltonian\cite{alan4}. This result should be contrasted with perturbation
theory about the free fermion Hamiltonian, which contains $(\log)^2$
singularities resulting from the vanishing Fermi velocity at the corners of the
square Fermi surface.\\
\\
Any non-zero interaction will lead to the irrelevancy of the Fermi velocity
modulation, but a much more restrictive condition results for the stability of
the square Fermi surface itself, away from half-filling. The appropriate
perturbation is the number operator which is given for the ``1'' face by:
\widetext
\begin{equation}
\displaystyle{\Delta \mu N_{1\sigma} = \Delta\mu (\pi k_F^{-1} ) \int dx
\sum_{l,l_1 l_2} ~g(l-l_1) g(l-l_2) \psi^\dagger_{1l_1 \sigma} (x)
\psi_{1l_2\sigma} (x) }
\end{equation}

\noindent
where $\Delta \mu$ is the change in chemical potential from half-filling. The
corresponding correlation function to second order is proportional to:
\begin{displaymath}
\sum_{l, l_1 , l_2} g^2(l-l_1) g^2(l - l_2)
\frac{1}{(R - V^\prime \tau + i \alpha)^2}
\left(\frac{\alpha^2}{R^2 - V^{\prime^2}\tau^2 +
\alpha^2}\right)^{\displaystyle{d (l_1 - l_2)}}
\end{displaymath}
\narrowtext

\noindent
since the two derivative operators are absent. For this to be irrelevant, $d
(l_1 - l_2) > 2$ for all $l_1 - l_2$, as discussed above. This requires a
sufficiently strong interaction to be present. Under this circumstance, the
only contribution is the diagonal term, $l_1 = l_2$, given by
\begin{equation}
N_{1\sigma} (x) = \frac{\pi}{2 k_F} \sum_{l_1} \rho_{1l,\sigma} (x)
\end{equation}

\noindent
where the factor of $1/2$ comes from $\sum_l ~ g^2 (l - l_1)$. The presence of
this term simply shifts the Fermi momentum uniformly, the square Fermi surface
remains a square.\\
\\
Strong interactions tend to localize single electrons to a chain, at least in
lowest order. In higher orders, or when interactions between adjacent faces
are included, it is possible that the situation might change. It seems
intuitively plausible that a crossover from a square to a rounded Fermi surface
 should occur, but it is not yet clear which terms are responsible for this.
Intra-facial hopping terms, such as $\psi_1^\dagger \psi_3 ,$ etc, would lead
to rounding. These do not appear in lowest order, but could enter in connection
with some of the relevant operators discussed in the following section. This
problem requires further attention. It is nonetheless very interesting that the
square Fermi surface is stable at this level.

\section{RELEVANT INTERACTIONS}

\noindent
Interactions between chains are generated through the mapping of the fermion
operators for the four faces onto independent chains. As described in the
previous section, it is possible to extract a purely bosonic interaction, and
use this to define a new starting point for perturbation theory. Other
interactions remain, and it is necessary to construct a new interaction
Hamiltonian, and attempt to solve this new problem. This situation is similar
to the Kondo problem or the backward scattering model, where a similar
reconstruction of the interactions is possible, leading to a new expansion.\\
\\
Consider the interactions involving parallel faces, the ``1'' and ``2''
operators. The lattice term $\displaystyle{U \sum_i n_{i +} n_{i -}}$ would
contain

\begin{equation}
\sum_i n_{i+} n_{i-} \rightarrow \sum_l \int ~dx~
\Psi^{\dagger}_{l+} (x) \Psi_{l,+} (x)
\Psi^{\dagger}_{1,-} (x) \Psi_{l,-} (x)
\label{eq:501}
\end{equation}

\noindent
with the $\Psi_{l, \sigma} (x) = \psi^\prime_{1l\sigma} (x) +
\psi^\prime_{2l\sigma} (x) $, and these primed operators are, in turn, sums
over independent chains given by Eq.\ (\ref{eq:206}).
There are similar terms for the ``3'' and ``4'' faces, and cross terms
containing interactions between adjacent faces. Cross terms are not considered
here. The task now is to extract the purely boson interaction terms in the sum,
giving $H_B$ of the previous sections, and find the remaining operators. These
new operators then provide the desired perturbation expansion.\\
\\
A conclusion of the previous section  is that the sum over independent chain
Fermi operators must have chain indices paired. If a single unpaired index
appears, that operator will contribute an operator dimension greater than
$1/2$, as in
Eq.\ (\ref{eq:413}), and the remaining three greater than $3/2$, giving an
overall dimension greater than $2$. In the second order correlation function
using
Eq.\ (\ref{eq:501}), the overall dimension will be greater than $4$, and hence
is an irrelevant term. This is another manifestation of the irrelevance of
single particle hopping for the square Fermi surface.\\
\\
Consider first the terms which occur when fermion operators with the same spin
are paired on the same chain. If they  have the same face index, there are
boson operators already included in $H_B$. Otherwise, there are two types of
terms
\widetext
\begin{equation}
H_{BS} = U \pi^2 k_F^{-2} \int ~dx~ \sum_{l_1 l_2} G (l_1 - l_2)
\left[\psi^\dagger_{1l_1 +} (x) \psi_{2l_1 +}^{} (x) \psi^\dagger_{2l_2 -} (x)
\psi_{1l_2 -}^{} (x) + \mbox{h.c.}\right]
\label{eq:502}
\end{equation}
\begin{equation}
H_U = \displaystyle{U  \pi^2 k_F^{-2}\int dx \sum_{l_1 l_2}
G(l_1 - l_2) \left[ \psi_{1l_1+}^\dagger (x)\psi_{2l_1 +}^{}(x)
\psi_{1 l_2 -}^\dagger (x) \psi_{2 l_2 -}^{} (x) + \mbox{h.c.}\right] }
\label{eq:503}
\end{equation}
where
\begin{displaymath}
G (l_1 - l_2) = \displaystyle{\sum_l g^2 (l - l_1) g^2 (l - l_2) =
(12)^{-1} \delta_{l_1 , l_2} +  2^{-1} \pi^{-2} (l_1 - l_2)^{-2} }.
\end{displaymath}

\noindent
Physically, $H_{BS}$ describes backward scattering on same or different chains,
while $H_U$ is an Umklapp process. \\
\\
In a 1D system, it is known that $H_U$ is relevant and $H_{BS}$ is
irrelevant for repulsive interactions\cite{alan3}, while these roles are
reversed for attractive interactions. The situation is similar here, as can be
shown using the fermion-boson duality principle from the previous section.\\
\\
The Umklapp term for $l_1 = l_2$ in the boson representation is given by:
\begin{equation}
\displaystyle{H_U =  -U G (0) (2 k_F \alpha)^{-2}  \sum_{l}
 \int ~dx  \left[e^{\displaystyle{-4i k_F x}} ~  e^{-\displaystyle{\sqrt{2}
{}~(\varphi_{1\rho l}^{} (x) + \varphi_{2 \rho l}^{} (x) ) }}
\mbox{+hc}\right]}
\label{eq:504}
\end{equation}
\narrowtext

\noindent
where the $\varphi_{1\rho l}$ operators are defined in
Eq.\ (\ref{eq:406}). $H_{BS}$ is the same with the $\varphi_{1\rho l}$
operators replaced by $\varphi_{1\sigma l}$ operators, and the $e^{4 k_F x}$
phase factor is absent. The minus sign results from ordering Eq.\
(\ref{eq:503}) into the form
$(-) \psi^\dagger_{1l +} \psi_{2 l -}^{} \psi_{1l -}^\dagger \psi_{2 l +}^{}$.
It can then be confirmed that the matrix element of this operator between the
states
$\langle\psi_{2l_1 -}^\dagger (x_2)  \psi_{1l_1 +}(x_2) | ~\mbox{and}~
\psi_{2l_1 +}^\dagger (x_1)  \psi_{1l_1 -} (x_1)\rangle$
is the same as the boson equivalent in Eq.\ (\ref{eq:504}), sandwiched between
the boson equivalents of these two states.\\
\\
Diagonalizing $H_o + H_B$ as in section IV with $S_1$, will transform the
exponent in Eq.\ (\ref{eq:504}) according to
\begin{equation}
\begin{array}{ll}
e^{S_1} & \displaystyle{ (2N)^{-1} \sum_{\vec{n}} n_x^{-1}  ~\left(\rho_1
(\vec{n}) +
\rho_2 (\vec{n})\right) ~
e^{-i k_x x-i p_y l}~e^{-S_1}}\\
&\\
&\displaystyle{= (2N)^{-1}  \sum_{\vec{n}} ~n^{-1}_x e^{-\eta (n_y)}
\left(\rho_1 (\vec{n}) + \rho_2 (\vec{n})\right) ~
e^{-i k_x x-i p_y l}}
\end{array}
\label{eq:505}
\end{equation}

\noindent
from which the scaling dimension of the Umklapp operator,  $d_U$ is seen to be
\begin{equation}
\displaystyle{d_U = \frac{1}{2N}\sum^N_{n_y - N} \left(2e^{-2\eta (n_y)}\right)
}
\label{eq:506}
\end{equation}

\noindent
as discussed in the previous section. This means the correlation function
appearing in second order using Eq.\ (\ref{eq:503}) falls off as
(distance)$^{-2 d_U}$. The zero interaction case, $\eta = 0$, has $2 d_U = 4$
and
for repulsive interactions $d_U < 2$ from Eq.\ (\ref{eq:405}). This term is
then relevant and must be retained. \\
\\
The phase factor $e^{4 i k_F k}$ for half-filling is evaluated to find $4 k_F x
= 4 (\pi 2^{-1/2} ~s^{-1} ) (s ~ 2^{-1/2}) (m + n ) = 0~
 (\mbox{mod}~ 2\pi)$ and makes no contribution. Notice that for strong
interactions, such that the Fermi surface remains square as discussed in the
previous section $k_F$ shifts away from $\pi s^{-1} 2^{-\frac{1}{2}}$ under
doping and the Umklapp term describes a (quantum) commensurate-incommensurate
transition\cite{alan15}. This transition will differ somewhat from the one
dimensional problem due to the parallel momentum dependence contained in
$H_B$.\\
\\
Using the diagonalizing transformation for the spin degrees of freedom
contained in Eq.\ (\ref{eq:502}), the operator dimension for equal chain
backward scattering is given by
\begin{equation}
\displaystyle{d_{BS} = \frac{1}{2N}\sum^N_{n_y = - N} ~\left(2 e^{-2\mu
(n_y)}\right)}
\label{eq:507}
\end{equation}

\noindent
and Eq.\ (\ref{eq:405}) shows $d_{BS} > 2$ for repulsive interactions. For
attractive  interactions, the roles are reversed, $d_U > 2$ and $d_{BS} < 2$.\\
\\
When $l_1 \neq l_2$ in Eq.\ (\ref{eq:503}), the scaling dimension is changed,
becoming dependent on $(l_1 - l_2)$ through the $k_y$ dependence in $H_B$. In
addition, the spin degrees of freedom do not cancel in the exponent when the
fermion operators are bosonized and contribute to the exponent of the power
law. The result for the operator scaling dimension is found to be:
\begin{eqnarray}
\displaystyle{d_U (l_1 - l_2) = \frac{1}{2N}~ \sum^N_{n_y = -N} \Bigl[e^{-2\eta
(n_y)}
\left(1 + e^{i p_y (l_1 - l_2)}\right)}\nonumber\\
\displaystyle{
+ e^{-2\mu (n_y)}
(1 - e^{i p_y (l_1 - l_2)}\Bigr]}
\label{eq:508}
\end{eqnarray}

\noindent
For repulsive interactions, $\eta > 0$ and $\mu < 0$, and the value of $d (l_1
- l_2)$ is not necessarily small. The result for $d_{BS} (l_1 - l_2)$ is found
from Eq.\ (\ref{eq:507}) by interchanging $\eta$ and $\mu$.\\
\\
At this stage of the analysis, it is helpful to reflect on the significance
of the transformations which change the operator dimensions, as determined by
$\eta$ and $\mu$. For general repulsive interactions, $\eta >  0$ and $\mu <
0$, from Eq.\ (\ref{eq:405}). For these signs, and sufficiently large
magnitudes,
$d_{BS} (l_1 - l_2) > 2$ for all $l_1 - l_2$, and $d_U (l_1 - l_2) > 2$ for
$l_1 \neq l_2$. These terms are irrelevant. The surviving relevant operator
from this pairing is given by chain diagonal Umklapp processes, with dimension
$d_U < 2$ from Eq.\ (\ref{eq:506}).\\
\\
With the coupling constants chosen to lie in this range, it is possible to
anticipate the existence of a solution. This will be constructed by dropping
the irrelevant operators, and expanding the relevant operators in the small
parameters, $e^{-2\eta}$ or $e^{2\mu}$. There is a similarity here to the 1D
massive Thirring model, where expanding the mass term when  the analogous
parameter is small, leads to the Klein-Gordon equation\cite{alan10}.\\
\\
It is necessary to examine other pairings of Eq.\ (\ref{eq:501})
to extract further relevant operators. Pairings which involve the first and
third terms (with second and fourth also paired) contain terms like
\\$\psi^\dagger_{1 l_1 +} \psi^\dagger_{2 l_1 -} \psi_{2 l_2 -}^{} \psi_{1 l_2
+}^{} ,  \psi^\dagger_{1 l_1 +} \psi^\dagger_{2 l_1 -} \psi_{1 l_2 -}^{}
\psi_{2 l_2 +}^{}$, and $\psi^\dagger_{1 l_1 +} \psi^\dagger_{1 l_1 -} \psi_{2
l_2 -}^{} \psi_{2 l_2 +}^{}$. After bosonizing these, it is found that they are
all irrelevant. The first two give the operator dimension of Eq.\
(\ref{eq:508}),
but with $\eta \rightarrow - \eta$ and $\mu \rightarrow - \mu$, and these are
larger than two in the coupling constant region of interest. The third
resembles single particle hopping with an extra $\sqrt{2}$ in the exponent
after bosonization, and is irrelevant.\\
\\
Cross terms which leave a factor of $e^{2 i k_F x}$ after bosonization can be
neglected since they have a momentum transfer of $2 k_F$. This would require a
large energy transfer and the intermediate states would lie far away from the
Fermi surface.\\
\\
The remaining pairing, with the first and fourth terms on the same chain
together with the second and third, contains relevant operators. These are
given by $-\psi^\dagger_{1 l_1 +} \psi_{2 l_1 -}^{} \psi^\dagger_{1 l_2 -}
\psi_{2 l_2 +}^{} $ and $-\psi^\dagger_{1 l_1 +} \psi_{2 l_1 -}^{}
\psi^\dagger_{2 l_2 -} \psi_{1 l_2 +}^{}$, plus the $(1\leftrightarrow 2 )$
interchange. Terms such as $\psi^\dagger_{1 l_1 +} \psi_{1 l_1 -}^{}
\psi^\dagger_{2 l_2 -} \psi_{2 l_2 +}^{}$ are irrelevant, and those involving
$e^{2 i k_F x}$ after bosonization can be discarded as previously discussed. \\
\\
Both relevant operators involve products of $2 k_F$ spin-flip excitations on
different chains. (The $l_1 = l_2$ terms has already been counted in the first
pairing, characterized by $d_U$  , $d_{BS}$ and a boson term). After
bosonizing, these terms
give respectively:
\widetext
\begin{equation}
\displaystyle{H_{DU} = - U (2 k_F \alpha)^{-2} \int ~ dx \sum_{l_1 \neq l_2} G
(l_1 - l_2) ~ \left\{e^{\displaystyle{ \varphi_{\rho l_1} (x) +\varphi_{\rho
l_2} (x)  + \tilde{\varphi}_{\sigma l_1} (x)  - \tilde{\varphi}_{\sigma l_2}
(x)}} + (1 \leftrightarrow 2)  \right\}}
\label{eq:509}
\end{equation}
\begin{equation}
\displaystyle{ H_{DB} = - U (2 k_F \alpha)^{-2} \int ~ dx \sum_{l_1 \neq l_2} G
(l_1 - l_2) ~ \left\{e^{\displaystyle{ \varphi_{\rho l_1} (x) -\varphi_{\rho
l_2} (x) + \tilde{\varphi}_{\sigma l_1} (x) -\tilde{\varphi}_{\sigma l_2} (x)
}} + (1 \leftrightarrow 2)\right\}}
\label{eq:510}
\end{equation}
\narrowtext

\noindent
Here, $\sqrt{2} \varphi_{\rho l} (x) = \varphi_{1\rho l} (x) + \varphi_{2 \cdot
\rho l}$ and $\sqrt{2} \tilde{\varphi}_{\sigma l} = \varphi_{1 \sigma l} (x) -
\varphi_{2 \sigma l} (x) $ with fields $\varphi_{1\sigma l} (x), $ etc, defined
in Eq.\ (\ref{eq:406}). It should be noted that $\varphi_{\rho l} (x)$ involves
the combination $\left[\rho_1  (\vec{n}) + \rho_2 (\vec{n})\right]$ which is
multiplied by $e^{-\eta (n_y)}$ under diagonalization, as in Eq.\
(\ref{eq:505}). Similarly, the $\tilde{\varphi}_{\sigma l} (x)$ involves
$\left(\sigma_1 (\vec{n}) - \sigma_2 (\vec{n})\right) $ which is multiplied by
$e^{\mu (n_y)}$ under diagonalization, from Eq.\ (\ref{eq:505}). Both
multiplicative factors should be small for the corresponding operator to be
relevant. Note that the fermion $(1 \leftrightarrow 2)$ interchange requires
$\varphi_{\rho l}$ to change sign, since there is a sign difference in the
boson representation of $\psi_1$ and $\psi_2$, while $\tilde{\varphi}_{\sigma
l}$ does not change sign. Under hermitian  conjugation both change sign.\\
\\
It is appropriate to label the first of these terms as ``different chain
Umklapp'', $H_{DU}$, while the second is a ``different chain backward
scattering'', $H_{DB}$. Their scaling dimensions, after diagonalizing $H_B$,
are given by
\begin{eqnarray}
d_{DU} (l_{12}) = \frac{1}{2N} \sum^N_{n_y =- N} \Bigl[e^{-2 \eta (n_y)}
\left(1 + e^{i p_y l_{12}}\right)\nonumber\\
 + e^{2\mu (n_y)} \left(1 -e^{i p_y l_{12}}\right) \Bigr]
\end{eqnarray}
\begin{eqnarray}
d_{DB} (l_{12}) = \frac{1}{2N} \sum^N_{n_y = - N}\Bigl[e^{-2 \eta (n_y)}
\left(1 - e^{i p_y l_{12}}\right)\nonumber\\
 + e^{2\mu (n_y)} \left(1 -e^{i p_y l_{12}}\right) \Bigr]
\end{eqnarray}

\noindent
Both are relevant (dimension less than two) for $\eta >0$ and $\mu < 0$, the
repulsive case of interest. There is a region of parameter space where all
three relevant operators have a small dimension. In this case, the exponents
contained in Eq.\ (\ref{eq:504}), Eq.\ (\ref{eq:509}), and Eq.\ (\ref{eq:510})
can be expanded.\\
\\
Scattering between parallel faces results in an equivalence to interacting
fermions on separate chains. In this section, the interaction terms have been
extracted and their scaling dimensions calculated. There is a strong coupling
region, characterized by three operators with small dimension, $d_U , d_{BU}$
and $d_{DB}$, which must be included. The following section presents the
construction of the solution.\\
\\
\section{GAP GENERATION AND THE INSULATING PHASE}

\noindent
This section begins with the boson problem of section III, adding the terms
with relevant operators from the previous section. It would be interesting to
find a solution for this problem, however this has not been found for general
coupling constants. But there is a limit where a solution is possible for
particular choices of coupling constants. This choice corresponds to the
parameters $e^{-2 \eta (n_y)}$ and $e^{2 \mu (n_y) }$ small, such that the
exponents of
Eq.\ (\ref{eq:504}), Eq.\ (\ref{eq:509}) and Eq.\ (\ref{eq:510})
can be expanded. Physically, this is a strong repulsion regime, with some
restrictions on the spin dependent interactions, $U_1$ and $U_2$ of section
III.\\
\\
The boson part of the problem, $H_o + H_B$ is given by Eq.\ (\ref{eq:401}).
{}From the preceding section, the three relevant operators must be added to
define a new ``interaction'' term. Diagonalizing $H_o + H_B$, using the $S$
operators of Eq.\ (\ref{eq:402}), transforms the ``phase'' fields appearing in
the relevant operators, generally decreasing them (for $e^{-\eta}$ and
$e^{\mu}$ both small). Expanding these phase factors to quadratic order gives
the new interaction Hamiltonian. For the 1-2 faces, this is:
\widetext
\begin{equation}
\begin{array}{lll}
H &=& H_D + H_{\mbox{int}}\\
H_{\mbox{int}} &=& \displaystyle{ -2 U (2 k_F \alpha)^{-2} \int ~ dx ~\sum_{l_1
, l_2}
G (l_1 - l_2)  e^S \left[ \varphi_{\rho l_1}^2 (x) + \varphi^2_{\rho l_2} (x) +
\left(\tilde{\varphi}_{\sigma l_1} - \tilde{\varphi}_{\sigma l_2} \right)^2
\right] e^{-S}}
\end{array}
\label{eq:601}
\end{equation}
\narrowtext

\noindent
where $H_D$ is Eq.\ (\ref{eq:403}) and  $S= S_1 + S_2$ from
Eq.\ (\ref{eq:402}). Not only do the first order terms cancel, but the cross
terms between spin and charge operators cancel as well. To this order, the
Hamiltonian separates into independent spin and charge sectors.\\
\\
After Fourier transforming, the new Hamiltonian can be written in terms of the
density operators as $ H = H_\rho + H_\sigma $, where:
\widetext
\begin{eqnarray}
H_{\rho} = \displaystyle{ \frac{\pi v_o}{2 NL } \sum_{\vec{n}}\Bigl\{
\gamma (n_y)  \left[ \rho_1  (\vec{n}) \rho_1 (-\vec{n}) +
\rho_2 (\vec{n}) \rho_2 (-\vec{n})\right]}\nonumber\\
\displaystyle{{}+ M^2_\rho (n_y) k_x^{-2} \left[
\rho_1 (\vec{n}) + \rho_2 (-\vec{n}) \right] \left[
\rho_1 (-\vec{n}) + \rho_2 (-\vec{n})\right]\Bigr\} }
\label{eq:602}
\end{eqnarray}
\begin{eqnarray}
H_\sigma = \displaystyle{ \frac{\pi v_o}{2 NL } \sum_{\vec{n}} \Bigl\{
\delta (n_y) \left[\sigma_1 (\vec{n}) \sigma_1 (-\vec{n}) + \sigma_2
(\vec{n})\sigma_2 (-\vec{n})\right]}\nonumber\\
\displaystyle{{}+ M^2_\sigma (n_y) k_x^{-2}
\left[\sigma_1 (\vec{n}) + \sigma_2 (\vec{n})\right]
\left[\sigma_1 (-\vec{n}) + \sigma_2 (-\vec{n})\right]\Bigr\} }
\label{eq:603}
\end{eqnarray}
\narrowtext

\noindent
with velocity renormalizations given by Eq.\ (\ref{eq:404}), and
\begin{equation}
\begin{array}{lll}
M^2_\rho (n_y) &=& \displaystyle{\frac{8\pi v_o U \gamma (n_y)}{(k_F
\alpha)^2}} ~ e^{\displaystyle{-2\eta (n_y)}} f^2 (0)\\
&&\\
M^2_\sigma (n_y) &=& \displaystyle{\frac{4\pi v_o U \delta (n_y)}{(k_F
\alpha)^2}}~ e^{\displaystyle{2 \mu (n_y)}} \left[f^2 (0) - f^2 (n_y)\right]\\
\end{array}
\end{equation}

\noindent
Here $\eta$ and $\mu$ are given by Eq.\ (\ref{eq:405}) and $f^2 (n_y)$ under
Eq.\ (\ref{eq:306}).  The combinations in these $M^2$ are chosen such that the
spectrum of Eq.\ (\ref{eq:602}) and Eq.\ (\ref{eq:603}) takes the
``relativistic'' form
\begin{equation}
\begin{array}{lll}
E^2_\rho (\vec{n}) &=& \displaystyle{k_x^2 v_o^2 \gamma^2 (n_y) + M^2_\rho
(n_y)}\\
&&\\
E^2_\sigma (\vec{n}) &=& \displaystyle{k_x^2 v_o^2 \delta^2 (n_y) + M^2_\rho
(n_y)}
\end{array}
\end{equation}

\noindent
as can easily be confirmed by diagonalizing these two. The contributions from
the 3-4 faces result in an identical Hamiltonian, with $\eta_x \leftrightarrow
n_y$.\\
\\
It is clear that the charge density gap is only weakly dependent on the
transverse momentum, through the velocity renormalization. On the other hand,
the spin density gap squared vanishes at $n_y = 0$, and increases linearly with
momentum. A quadratic dependence on $n_y$ would have been intuitively
understandable, since it could be renormalized to give an isotropic excitation
spectrum, proportional to $n^2_x + n_y^2$. Perhaps the inclusion of
interactions between adjacent faces could change $|n_y|$ to $n_y^2$, but that
remains unclear at present.\\
\\
Renormalization of the bare coupling constant $U$ in
Eq.\ (\ref{eq:601}) is required, if the limit $\alpha \rightarrow 0$ is to be
taken, as in the purely 1D problem\cite{alan8}. Alternatively, one can regard
the $M^2$ quantities, as physically observable, setting a length scale for
which the bare parameters are determined.\\
\\
This completes the present analysis of interacting electrons on a square Fermi
surface. It focussed on interactions between
parallel faces, although the method can be readily generalized to adjacent face
interactions. Perhaps the most interesting result concerns the mapping onto
chains, and the ability to extract relevant operators in the chain picture.\\
\\
It should be possible to study this mapping more systematically using
renormalization group scaling. If each non-boson operator is assigned a
coupling constant, as in 1D systems\cite{alan3}, these will satisfy scaling
equations. Presumably, some will scale to stronger coupling strengths, and
solutions such as those presented here, could characterize the asymptotic
behavior. It will be interesting to confirm, or disprove, the strong coupling
region outlined here.\\
\\
Doping away from this point will generate a phase factor in the Umklapp terms,
$H_U$ and $H_{DU}$, but not in $H_{DB}$. This phase factor of $\exp (4 i \delta
k_F x)$, where $\delta k_F $ is the departure from the half-filled value,
changes the nature of the problem. The low energy excitations then include some
charge  states, and it might be expected that a new type of generalized
Luttinger model results.\\
\\
Spin excitations are more problematic in this solution. Although the spin
excitation energy vanishes when $p_y \rightarrow 0$, it does not vanish
linearly. At $k_x = 0$, the spectrum is proportional to $|p_y |^{1/2}$. It is
not clear if this is an artifact of parallel face interactions and will be
modified by adjacent face scattering. Another possibility is contained in the
problem of uniform phase ordering in the chain mapping. As indicated under
Eq.\ (\ref{eq:407}), a uniform phase on each chain is permitted. Perhaps other
configurations can be found with lower ground state energy than the present
result.

\section{DISCUSSION AND CONCLUSION}

\noindent
This paper has presented a mapping of the square Fermi surface onto two sets of
chains, one set for each axis of the square. It correctly describes electronic
excitations near the half-filled band, with the average occupation number near
one. Interactions between electrons separate naturally into boson operators on
these chains, plus additional operators. The kinetic energy plus the boson
operators define a new starting point for perturbation theory, with the
additional operators as the perturbation.\\
\\
This method provides a very powerful new tool for analyzing the properties of
the square Fermi surface. The purely bosonic part of the problem can be
diagonalized exactly, giving behavior similar to the Luttinger model. It
exhibits power low correlations functions and has no jump in the occupation
number  at the Fermi level. Modulation of the Fermi velocity along the face of
the square is irrelevant. Rounding of the square is similarly irrelevant for
sufficiently strong interaction strengths.\\
\\
All of these results are derived using the new perturbation theory about the
boson part of the problem. The interactions also contain non-boson terms, which
have been analyzed using boson-fermion duality for the chain representations.
There is a region of strong coupling when the relevant operator generated by
interactions can be solved exactly, in the sense that the sine-Gordon problem
becomes the
Klein-Gordon problem for strong fermion interactions\cite{alan10}. For
half-filling, a gap appears in the charge spectrum, while the spin excitations
remain gapless.\\
\\
While the mapping works for very general interactions, this paper has
concentrated on a special case of parallel face interactions. For a complete
solution, interactions between adjacent faces and possible momentum dependences
need to be included. A convincing solution of the Hubbard model awaits an
analysis of these generalizations.\\
\\
When doped away from half-filling, spinless fermion states above the charge gap
are occupied. This situation is reminiscent of the commensurate-incommensurate
transition\cite{alan15}, with additional complications due to the transverse
momentum dependence. Although the spin degrees of freedom are unchanged, it
will be interesting to explore the changes in the charge excitations. In this
way, the very interesting issue of superconducting correlations can be
resolved.\\
\\
Lattice degrees of freedom must also be included, and a logical procedure would
involve diagonalizing the interesting electron problem first. Adding in the
electron-phonon coupling, selecting out the relevant operators, and solving
this new problem, would cast new light on the problem of superconductivity in
the Cuprate compounds. It seems these steps are all possible within the mapping
framework.\\
\\
It would also seem this problem of interacting fermions on a square Fermi
surface could be addressed in other ways. The requirement that fermions must be
paired on separate chains could be realized by a gauge symmetry. Since the
single chain problem can be solved by a Bethe Ansatz\cite{alan16}, perhaps this
new model of coupled chains has a solution involving superpositions of these
single chain Bethe Ansatz hypothesis wavefunctions. It might be simpler to test
this on the spinless fermion problem. \\
\\
Solutions for the problem of interacting electrons on a square Fermi surface,
of the type presented here, represent a new regime with correlation effects
dominant. Much of the intuition based on Fermi liquid behavior, is no longer
applicable. If it can be shown that the high temperature superconductors fall
in this regime, this is truly strongly correlated superconductivity.

\end{document}